\documentclass[conference]{IEEEtran}
\usepackage{amsmath,amsthm}
\usepackage{cite}
\usepackage{graphicx}
\usepackage{epstopdf}
\usepackage{amsfonts,amsmath,amssymb}
\usepackage{cite}
\usepackage{graphicx}
\usepackage{url}
\usepackage{bm}
\usepackage{bbm}
\usepackage{amssymb}

\begin{document}

\title{{Entanglement Generation via Non-Gaussian Transfer over Atmospheric Fading Channels}}
\author{
\IEEEauthorblockN{Nedasadat Hosseinidehaj, Robert Malaney}
\IEEEauthorblockA{School of Electrical Engineering  \& Telecommunications,\\
The University of New South Wales,\\
Sydney, NSW 2052, Australia\\
neda.hosseini@unsw.edu.au, r.malaney@unsw.edu.au}
}

\vspace{-5cm}

\maketitle
\begin{abstract}

In this work we probe the usefulness of non-Gaussian entangled states as a resource for quantum communication through atmospheric channels. We outline the initial conditions in which non-Gaussian state transfer leads to enhanced entanglement transfer relative to that obtainable via Gaussian state transfer. However, we conclude  that in (anticipated) operational scenarios - where most of the non-Gaussian states to be transferred over the air are created just-in-time via photonic subtraction, addition or replacement from incoming Gaussian states - the entanglement-generation rate between stations via non-Gaussian state transfer will be substantially less than that created by direct Gaussian state transfer.  The role of  post-selection, distillation and quantum memory in altering this conclusion is discussed, and comparison with entanglement rates produced via single-photon technologies is provided. Our results suggest that in the near term  entangled Gaussian states, squeezed beyond some modest level, offer the most attractive proposition for the distribution of entanglement through high-loss atmospheric channels. The implications of our results for entanglement-based QKD to low-earth orbit are presented.

\end{abstract}

\section{Introduction}
 The deployment of systems that provide for the distribution of entangled quantum states via satellite would represent an important step in the pursuit of a global quantum communications network \cite{s1, s4, s5, s7, s10, s11}. However, a serious issue that will be faced by such systems is the unavoidable degradation of entanglement caused by atmospheric effects, most notably atmospheric turbulence \cite{fso}. From the perspective of future quantum communications, it is therefore important to fully quantify this anticipated entanglement degradation, and to pursue system designs that minimize it.

In the continuous-variable (CV) space, previous efforts in this regard have largely focussed on  the transmittance of  Gaussian entangled states through atmospheric channels \cite{Dong, Usenko, Heim, 1}. Although Gaussian quantum states are a well-established resource from both a theoretical and an experimental perspective (for review see\cite{Gaussian}), the use of CV non-Gaussian quantum states as a means for quantum communication has also garnered interest \cite{1st_PSS, 2,1st_PAS, telep, 3, 4, Amp1, Amp2, 5, 6, ME, 8, 9, Oxford}. Consideration of non-Gaussian states is interesting for many reasons, including teleportation \cite{1st_PSS, 2, 1st_PAS, telep, 3, ME, 9, Oxford}, and cloning \cite{Amp1, Amp2}. However,  in the context of quantum communications via satellite, entangled non-Gaussian states are  particularly interesting for two key reasons. First, non-Gaussian states can undergo entanglement distillation without the requirement for further non-Gaussian operations or de-Gaussification procedures  - an outcome forbidden for Gaussian states \cite{no_go1,no_go2}. Second,  in some circumstances, the entanglement intrinsic to non-Gaussian states is more robust against decoherence compared to the entanglement intrinsic to Gaussian states \cite{6, ME, 8}.

Previous works \cite{4, Amp1, Amp2, 5, 6, ME, 8} on the robustness of non-Gaussian entanglement have focussed on noisy \emph{fixed} attenuation (or amplification) channels. However, illuminating as such studies are we note that the turbulent atmosphere between ground and low-earth orbit (LEO) leads to an attenuation channel that is highly \emph{stochastic} in nature (i.e. a fading channel). As such, in the context of entanglement transfer through atmospheric channels towards (and from) LEO, it remains unclear if non-Gaussian states actually represent an effective resource relative to Gaussian states. Further, in operational modes the non-deterministic nature of the  operations that produce many non-Gaussian states can have a dramatic effect on their perceived usefulness \cite{Oxford, Oxford2013}. For example, in the deployment  scenario we focus on in this work, most of the non-Gaussian entangled states will be created dynamically via  non-Gaussian operations  on input two-mode squeezed vacuum (TMSV) states that would otherwise be used directly in the communication channel.


It is the purpose of the present work to include both channel fading  and non-deterministic production effects in
a detailed comparison study of the ground-to-LEO (and vice-versa) entanglement-generation rates offered by a wide class of non-Gaussian entangled states. Our initial focus will be the use of such states in \emph{first-generation} deployments aimed at delivering real-time quantum keys between Earth and LEO satellites, without the assistance of any entanglement distillation, post-selection, or quantum memory. Later in the paper we will explore the impact such assistance can have on the outcomes of second-generation deployments. We will also discuss our results in the context of the most well-known non-Gaussian entangled state - single photons combined into a Bell pair.

\begin{figure}[!t]
    \begin{center}
   {\includegraphics[width=3.5 in, height=2.5 in]{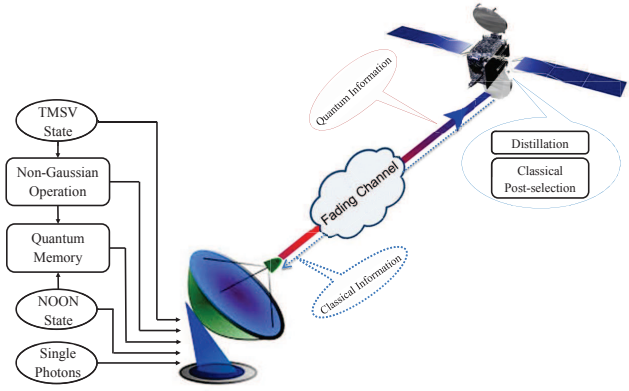}}
    \caption{An example of our system model in the  ground-to-LEO configuration.}\label{fig:0a}
    \end{center}
\end{figure}

The  remainder of this paper is as follows. In Sec.~II we detail our system model,  and describe the  evolution of our entangled states over the atmospheric channel. In Sec.~III we present our key results. In Sec. IV we discuss  the potential impact of distillation, post-selection, and quantum memory; and provide a comparison with single-photon technologies in the context of quantum key distribution (QKD).

\section {System Model and Quantum State Evolution}
We outline our system model, the quantum entangled states adopted, the evolution of the states through the atmospheric channel, and determination of the final entanglement.
\subsection {System Model}
In free-space channels the transmittance fluctuates due to atmospheric effects. Such fading channels can be characterized by a distribution of transmittance (transmission) factors $\eta $ with a probability density distribution $p\left( \eta  \right)$. Consistent with other recent studies \cite{fso, beamwander, Usenko}, we will assume that transmittance fading arising from the atmosphere is due only to beam wander. Assuming the beam spatially fluctuates around the center of the receiver's aperture, the probability density distribution $p\left( \eta  \right)$ can be described by the log-negative Weibull distribution \cite{beamwander},
\begin{equation}\
p\left( \eta  \right) = \frac{{2{L^2}}}{{\sigma _b^2\gamma_s \eta }}{\left( {2\ln \frac{{{\eta _0}}}{\eta }} \right)^{\left( {\frac{2}{\gamma_s }}- 1 \right) }}\exp \left( { - \frac{{{L^2}}}{{2\sigma _b^2}}{{\left( {2\ln \frac{{{\eta _0}}}{\eta }} \right)}^{\left( {\frac{2}{\gamma_s }} \right)}}} \right)
\label{f1}
\end{equation}
for $\eta  \in \left[ {0,\,{\eta _0}} \right]$, with $p\left( \eta  \right) = 0$ otherwise.
Here, $\sigma _b^2$ is the beam wander variance,
 $\gamma_s$ is the shape parameter,  $L$ is the scale parameter, and ${\eta _0}$ is the  maximum transmittance value. The latter three parameters are given by
\begin{eqnarray}\label{f2}
\begin{array}{l}
\gamma_s  = 8h\frac{{\exp \left( { - 4h} \right){I_1}\left[ {4h} \right]}}{{1 - \exp \left( { - 4h} \right){I_0}\left[ {4h} \right]}}{\left[ {\ln \left( {\frac{{2\eta _0^2}}{{1 - \exp \left( { - 4h} \right){I_0}\left[ {4h} \right]}}} \right)} \right]^{ - 1}}\\
\\
L = \beta{\left[ {\ln \left( {\frac{{2\eta _0^2}}{{1 - \exp \left( { - 4h} \right){I_0}\left[ {4h} \right]}}} \right)} \right]^{ - \left( {{1 \mathord{\left/
 {\vphantom {1 \gamma_s }} \right.
 \kern-\nulldelimiterspace} \gamma_s }} \right)}}\\
\\
\eta _0^2 = 1 - \exp \left( { - 2h} \right) ,
\end{array}
\end{eqnarray}
where ${I_0}\left[ . \right]$ and ${I_1}\left[ . \right]$ are the modified Bessel functions, and where $h = {\left( {{\beta \mathord{\left/
 {\vphantom {a W}} \right.
 \kern-\nulldelimiterspace} W}} \right)^2}$, with $\beta$ being the receiver aperture radius and $W$ the beam-spot radius. In our subsequent calculations we will adopt $W=1.1\beta$, and let the mean fading loss be controlled only by adjustments to the value of $\sigma _b$.

Since depolarization is very weak in the atmospheric channel, dephasing  will also be weak and thus we will ignore it \cite{sem}. We will initially consider the ensemble-average state, i.e. the case where the specific realization of the channel (transmittance value) is unknown. (The passing a local oscillator through the channel in an orthogonal polarized mode to the signal and measuring it in real-time at the receiver will be discussed later in the paper.)
We will also assume the sending station initially possesses a two-mode (mode~1 and mode~2) entangled state, with one (or more) of the modes transmitted to the receiving station(s) through the atmospheric channels whose fading characteristics are as described above. This leads to two operational settings.

\emph{ Asymmetric Setting}. In this setting we will assume one mode, mode~1, remains at the ground station (satellite), while the other mode, mode~2, is transmitted to the satellite (ground station) over the fading uplink (downlink) with probability density distribution ${p\left( \eta  \right)}$. A schematic illustration of this setting in an example uplink configuration is shown in Fig.~\ref{fig:0a}.
 The density operator of the two-mode state at the ground station and satellite for each realization of the transmission factor $\eta $ is given by ${\rho }(\eta )$. Since $\eta $ is a random variable, the elements of the total density operator of the resulting mixed state ${\rho^t}$ are calculated by averaging the elements of density operator ${\rho}(\eta )$ over all possible transmission factors of the fading channel giving
$\rho ^t = \int_0^{{\eta _0}} {p\left( \eta  \right){\rho }\left( \eta  \right)\,d\eta }$.

\emph{Symmetric Setting.} In this setting  we will assume the satellite initially possesses a two-mode entangled state, with one mode, mode~1, transmitted to the ground station 1 over a fading downlink with probability density distribution ${p_1}({\eta _1})$, while the other mode, mode~2, is transmitted to the ground station 2 over a different fading downlink with probability density distribution ${p_2}({\eta _2})$. Here, the two fading downlinks are assumed to be independent and identically distributed. The density operator of the two-mode state at the ground stations for each realization of the transmission factors ${\eta _1}$ and ${\eta _2}$ is given by the density operator ${\rho}({\eta _1},{\eta _2})$. The elements of the total density operator of the resulting mixed state ${\rho^t}$ are calculated by averaging the elements of density operator ${\rho }({\eta _1},{\eta _2})$ over all possible transmission factors of the two separate fading channels giving
$\rho ^t = \int_0^{{\eta _{0}}} {\int_0^{{\eta _{0}}} {{p_1}\left( {{\eta _1}} \right){p_2}\left( {{\eta _2}} \right){\rho }\left( {{\eta _1},{\eta _2}} \right)\,d{\eta _1}d{\eta _2}} }$.

Our two key performance indicators will be $E_{LN}$, the entanglement (logarithmic negativity); and ${R_E}$, the entanglement-generation rate. $E_{LN}$ represents the entanglement generated between two stations following the transfer of a pulse through the lossy fading channel(s);\footnote{When one mode is retained by the sender, the pulse refers to the second mode sent between the sender and receiver. When one mode is sent to one receiver and the other mode sent to a different receiver, the pulse refers to the two modes collectively.}
and ${R_E}$  encapsulates directly the probability of creating the initial state. Introducing ${P_c}$, the creation probability of the initial state (we adopt $P_c=1$ for TMSV states), we have $R_E = {P_c}\,E_{LN}$.
Note, ${R_E}$  is  in units of entanglement per initial pulse, where by initial pulse we mean the original TMSV pulse.\footnote{The entanglement rate  in units of entanglement/second can by calculated  as ${T_rR_E}$, where ${T_r}$ is the generation rate of TMSV states (per second) at the TMSV source.  If $P_c=1$ for a state then its rate into the channel is  $T_r$. However, in  our comparison tests the value of $T_r$ will not be important. Note that for NOON states, by initial pulse we mean the initial NOON state.}


The range of losses we consider cover a wide range of anticipated scenarios for
 a communication loop with LEO satellites. Such a loop which should be well covered by losses in the range 5dB (downlink) to 30dB losses (uplink), especially when we bear in mind the possibility of adaptive optic solutions being able to compensate significantly for beam wander (e.g. \cite{adapt}). Although not our focus here, our results will also be applicable to direct line-of-sight terrestrial communications through air.
\subsection{Entangled States}
For the Gaussian entangled state we adopt the TMSV state. This state is generated deterministically by non-linear optical processes (e.g. \cite{Gaussian}),
and is described in the Fock basis as
$\left| {TMSV} \right\rangle  = \sum\limits_{n = 0}^\infty  {{q_n}} {\left| n \right\rangle _1}{\left| n \right\rangle _2}$, where ${q_n} = {\lambda ^n}\sqrt {1 - {\lambda ^2}} $, and
where $\lambda  = \tanh r $, $r \in \mathbb{R}$ being a squeezing parameter, and indices 1 and  2 indicating the two modes. The two-mode squeezing in dB is given by $ - 10{\log _{10}}\left( {\exp ( - 2r)} \right)$.

For non-Gaussian entangled states we consider photon-subtracted squeezed (PSS) states \cite{1st_PSS, 2, telep, 3, 9, Oxford, Oxford2013, 7}, photon-added squeezed (PAS) states \cite{1st_PAS, telep, Oxford, 7, added}, photon-replaced squeezed (PRS) states \cite{ME, Oxford}, and NOON states \cite{NOON0, NOON1, NOON2}. Such states cover a wide range of the non-Gaussian state possibilities, and represent the  non-Gaussian states most likely to be used in future quantum application and communication  deployment scenarios.

For generation of a PSS state, we assume each mode of an incoming TMSV state interacts with a vacuum mode in a beam splitter with transmissivity $T$ (all beam splitters discussed here will have their transmissivity given by $T$). The two outputs of the beam splitter feed single-photon detectors. When both detectors register one photon simultaneously, a pure non-Gaussian state is heralded with probability $P_{sb}$. The resulting normalized state arising from this process and its creation probability are given by \cite{2, 3, Oxford}
\begin{eqnarray}\label{PSSb}
\begin{array}{l}
\left| {PS{S_b}} \right\rangle  = \sum\limits_{n = 0}^\infty  {{q_n}} {\left| n \right\rangle _1}{\left| n \right\rangle _2},\,{\rm where}\\
\\
{q_n} = \sqrt {\frac{{{{\left( {1 - {\lambda ^2}{T^4}} \right)}^3}}}{{1 + {\lambda ^2}{T^4}}}} {\left( {\lambda {T^2}} \right)^n}(n + 1)\\
\\
{P_{sb}} = \frac{{{\lambda ^2}\left( {1 - {\lambda ^2}} \right)\left( {1 + {\lambda ^2}{T^4}} \right){{\left( {1 - {T^2}} \right)}^2}}}{{{{\left( {1 - {\lambda ^2}{T^4}} \right)}^3}}}.
\end{array}
\end{eqnarray}
We will also study a PSS state where the photon subtraction as described above is applied to a single mode of the TMSV state (all single-mode non-Gaussian operations discussed here will apply to mode~2 of the original TMSV state). The normalized state arising from this process and its creation probability $P_{ss}$ are given by \cite{Oxford}
\begin{eqnarray}\label{PSSs}
\begin{array}{l}
\left| {PS{S_s}} \right\rangle  = \sum\limits_{n = 0}^\infty  {{q_n}} {\left| {n + 1} \right\rangle _1}{\left| n \right\rangle _2},\,{\rm where}\\
\\
{q_n} = \left( {1 - {\lambda ^2}{T^2}} \right){\left( {\lambda T} \right)^n}\sqrt {n + 1} \\
\\
{P_{ss}} = \frac{{{\lambda ^2}\left( {1 - {\lambda ^2}} \right)\left( {1 - {T^2}} \right)}}{{{{\left( {1 - {\lambda ^2}{T^2}} \right)}^2}}}.
\end{array}
\end{eqnarray}
Creation of non-Gaussian states via photon subtraction techniques as described above has been experimentally demonstrated \cite{exp1, exp2}.
Note that use of other kinds of photon detectors such as on/off detectors for obtaining a PSS state from a TMSV state has been studied in \cite{3}, but we will not consider such production here.

Addition of single photons to coherent states and thermal states of light has been experimentally realized \cite{added_exp1, added_exp2}.
For generation of our PAS state, we assume a single photon is added to each mode of a TMSV state  at a beam splitter, with the outputs of the beam splitter entering photon detectors. When a vacuum state is registered in both detectors simultaneously a pure non-Gaussian state is obtained with probability $P_{ab}$. The resulting normalized state and its creation probability are given by \cite{added, Oxford}
\begin{eqnarray}\label{PASb}
\begin{array}{l}
\left| {PA{S_b}} \right\rangle  = \sum\limits_{n = 0}^\infty  {{q_n}} {\left| n \right\rangle _1}{\left| n \right\rangle _2},\,{\rm where}\\
\\
{q_n} = \sqrt {\frac{{{{\left( {1 - {\lambda ^2}{T^4}} \right)}^3}}}{{1 + {\lambda ^2}{T^4}}}} {\left( {\lambda {T^2}} \right)^{n - 1}}n\\
\\
{P_{ab}} = \frac{{\left( {1 - {\lambda ^2}} \right)\left( {1 + {\lambda ^2}{T^4}} \right){{\left( {1 - {T^2}} \right)}^2}}}{{{{\left( {1 - {\lambda ^2}{T^4}} \right)}^3}}}.
\end{array}
\end{eqnarray}
A normalized PAS state obtained by applying the photon addition as described above to a single mode of the TMSV state, and its creation probability $P_{as}$ are given by \cite{Oxford}
\begin{eqnarray}\label{PASs}
\begin{array}{l}
\left| {PA{S_s}} \right\rangle  = \sum\limits_{n = 0}^\infty  {{q_n}} {\left| n \right\rangle _1}{\left| {n + 1} \right\rangle _2},\,{\rm where}\\
\\
{q_n} = \left( {1 - {\lambda ^2}{T^2}} \right){\left( {\lambda T} \right)^n}\sqrt {n + 1} \\
\\
{P_{as}} = \frac{{\left( {1 - {\lambda ^2}} \right)\left( {1 - {T^2}} \right)}}{{{{\left( {1 - {\lambda ^2}{T^2}} \right)}^2}}}.
\end{array}
\end{eqnarray}
Note also that in the process of photon addition the final probability of generating a PAS state is obtained by multiplying $P_{as}$ ($P_{ab}$) by the production probability of the required one (two) additional photon(s). Here, we will assume that single photons can be created with probability one.\footnote{At the present time there exists no experimental technique to achieve this, and therefore we can currently consider the entanglement rates calculated under this assumption to be limited to upper limits. This same assumption (and limitation) also applies to our discussion of PRS and NOON states.}


For generation of a PRS state, we assume each mode of a TMSV state   interacts with a single photon in a beam splitter, with the outputs of the beam splitter entering photon detectors. When both detectors register one photon simultaneously, a pure non-Gaussian state is heralded with probability $P_{rb}$. The resulting normalized state and its creation probability are given by \cite{Oxford}\footnote{A typographical error in the $P_{rb}$ of \cite{Oxford} is corrected for here.}
\begin{eqnarray}\label{PRSb}
\begin{array}{l}
\left| {PR{S_b}} \right\rangle  = \sum\limits_{n = 0}^\infty  {{q_n}} {\left| n \right\rangle _1}{\left| n \right\rangle _2},\,{\rm where}\\
\\
{q_n} = \frac{1}{{\sqrt {{P_{rb}}} }}\sqrt {1 - {\lambda ^2}} {\lambda ^n}{T^{2\left( {n - 1} \right)}}{\left( {{T^2} - n\left( {1 - {T^2}} \right)} \right)^2}\\
\\
{P_{rb}} = \frac{{1 - {\lambda ^2}}}{{{{\left( {1 - {T^4}{\lambda ^2}} \right)}^5}}}\\
\\
 \times \left[ {{T^4} + \left( {1 - 8{T^2} + 24{T^4} - 32{T^6} + 11{T^8}} \right){\lambda ^2}} \right.\\
\\
 + {T^4}\left( {11 - 56{T^2} + 96{T^4} - 56{T^6} + 11{T^8}} \right){\lambda ^4}\\
\\
 + {T^8}\left( {11 - 32{T^2} + 24{T^4} - 8{T^6} + {T^8}} \right){\lambda ^6} + {T^{12}}\left. {{\lambda ^8}} \right].
\end{array}
\end{eqnarray}
Considering operation of the photon replacement as discussed above on a single mode of the TMSV state, the resulting normalized state and its creation probability $P_{rs}$ are given by \cite{Oxford}
\begin{eqnarray}\label{PRSs}
\begin{array}{l}
\left| {PR{S_s}} \right\rangle  = \sum\limits_{n = 0}^\infty  {{q_n}} {\left| n \right\rangle _1}{\left| n \right\rangle _2}\,,{\rm where}\\
\\
{q_n} = \frac{1}{{\sqrt {{P_{rs}}} }}\sqrt {1 - {\lambda ^2}} {\lambda ^n}{T^{n - 1}}\left( {{T^2} - n\left( {1 - {T^2}} \right)} \right)\\
\\

{P_{rs}} = \frac{{1 - {\lambda ^2}}}{{{{\left( {1 - {T^2}{\lambda ^2}} \right)}^3}}}\\
\\
 \times \left[ {\left. {{\lambda ^2} + {T^2}\left( {1 + \left( {{T^2} - 4} \right){\lambda ^2} + {\lambda ^4}} \right)} \right]} \right..
\end{array}
\end{eqnarray}
The creation probability of a PSS state and a PAS state is reduced by increasing the transmissivity of the beam splitter, such that the creation probability is zero for $T=1$, while the creation probability of a PRS state is increased with transmissivity, such that the creation probability is one for $T=1$. We note that the resultant PRS state is identical to the original TMSV state for $T=1$.

NOON states are another form of well-known non-Gaussian states. Such states have been studied extensively in the context of quantum metrology where they can be used to obtain high-precision phase measurements (e.g. \cite{NOON2015}).
NOON states are described in the Fock basis as
$\left| {NOON} \right\rangle  = \frac{1}{{\sqrt 2 }}\left( {{{\left| n \right\rangle }_1}{{\left| 0 \right\rangle }_2} + {{\left| 0 \right\rangle }_1}{{\left| n \right\rangle }_2}} \right)$.
Using the interference of a single mode squeezed state and a classical coherent state on a 50:50 beam splitter, \cite{NOON_exp2,NOON_exp1} demonstrated the experimental production of NOON states up to $n=5$.
On the theoretical side, different proposals for NOON state generation have been considered (e.g. \cite{NOON0, NOON1, NOON2}). In \cite{NOON0} it is shown how a NOON state can be prepared from the vacuum state $\left| 0 \right\rangle \left| 0 \right\rangle $ by applying $n$ times a non-unitary transformation which is implemented probabilistically using only single-photon sources, linear optics, and photo-detectors. In this  scheme the optimum probability for generation of the entangled NOON state is given by
${P_n} = (n - 1)!{(2n)^{1 - n}}$,
and we will adopt this as the probability of  production for $n>2$ NOON states (we note two-photon NOON states can be created deterministically).\footnote{In this work we will not consider the single-particle entanglement represented by the single-photon NOON  state.}

\subsection{Evolution of Entangled States Over a Lossy Channel}

Unlike Gaussian states, non-Gaussian states are not completely characterized by the first and second moments of the quadrature operators. Therefore, we cannot quantify the evolution of non-Gaussian states solely through the covariance matrix. Previous works have looked at the evolution of non-Gaussian states through noisy loss channels through a Master equation approach \cite{4}, a characteristic function approach \cite{ME} or through a Kraus operator approach \cite{6}.

 We will employ the Kraus representation \cite{10} in order to directly analyze the action of the channel on our states. Considering a quantum state with density operator ${\rho _{in}}$ as the input of a trace-preserving completely positive channel, the output density operator of the channel can be described in an operator-sum representation of the form
${\rho _{out}} = \sum\limits_{\ell  = 0}^\infty  {{G_\ell }{\rho _{in}}\,G_\ell ^\dag }$,
where the Kraus operators ${G_\ell }$ satisfy $\sum\limits_{\ell  = 0}^\infty  {{G_\ell }\,G_\ell ^\dag }  = I$, with $I$ being the identity operator.
A noisy attenuator channel with the transmission factor $0 \le \eta  \le 1$ and an additional Gaussian noise $\chi  \ge 0$ can be realized by the composition of two `noiseless' channels  $\gamma_1$ and $\gamma_2$ via $\gamma_2(\phi)\circ \gamma_1(\zeta)$, where $\phi  = \sqrt {1 + \chi /2} $ and $\zeta  = \eta /\sqrt {1 + \chi /2} $.
The Kraus operators of $\gamma_2$ and $\gamma_1$ can be written respectively as,
\begin{equation}\
G_\ell ^{{\gamma _2}}\left( \phi  \right) = {\phi ^{ - 1}}\sum\limits_{m = 0}^\infty  {\sqrt {{}^{m + \ell }{C_\ell }} {{\left( {\sqrt {1 - {\phi ^{ - 2}}} } \right)}^\ell }{\phi ^{ - m}}\left| {m + \ell } \right\rangle \left\langle m \right|,}
\label{Amp}
\end{equation}
and
\begin{eqnarray}\label{Atten}
G_\ell ^{{\gamma _1}}\left( \zeta  \right) = \sum\limits_{m = 0}^\infty  {\sqrt {{}^{m + \ell }{C_\ell }} {{\left( {\sqrt {1 - {\zeta ^2}} } \right)}^\ell }{\zeta ^m}\left| m \right\rangle \left\langle {m + \ell } \right|,}
\end{eqnarray}
where ${}^{m + \ell }{C_\ell }$ is the binomial coefficient. From these operators it can then be shown that the action of the composite channel on the elementary density operator $\left| m \right\rangle \left\langle n \right|$ can be written as \cite{6},
\begin{eqnarray}\label{k2}
\begin{array}{l}
{\gamma _2}(\phi ) \circ {\gamma _1}(\zeta )\left( {\left| m \right\rangle \left\langle n \right|} \right) \to \\
\\
{\phi ^{ - 2}}\sum\limits_{\ell ' = 0}^\infty  {\sum\limits_{\ell  = 0}^{\min \{ m,n\} } {\sqrt {{}^{m - \ell  + \ell '}{C_{\ell '}}{}^{n - \ell  + \ell '}{C_{\ell '}}{}^m{C_\ell }{}^n{C_\ell }} } } \\
\\
 \times {\left( {{\phi ^{ - 1}}\zeta } \right)^{\left( {m + n - 2\ell } \right)}}{\left( {1 - {\phi ^{ - 2}}} \right)^{\ell '}}{\left( {1 - {\zeta ^2}} \right)^\ell }\\
\\
 \times \left| {m - \ell  + \ell '} \right\rangle \left\langle {n - \ell  + \ell '} \right|.
\end{array}
\end{eqnarray}

Let us consider a two-mode entangled state $\left| \psi  \right\rangle  = \sum\limits_{n = 0}^\infty  {{q_n}} {\left| n \right\rangle _1}{\left| n \right\rangle _2}$ for which the initial density operator can be writte,
\begin{eqnarray}\label{initial}
{\rho _{in}} = \sum\limits_{m = 0}^\infty  {\sum\limits_{n = 0}^\infty  {{q_m}{q_n}{{\left| m \right\rangle }_1}{{\left\langle n \right|}_1} \otimes {{\left| m \right\rangle }_2}{{\left\langle n \right|}_2}} } .
\end{eqnarray}
In the symmetric setting we assume one channel has transmittance $\eta={\eta_1}$ and  noise $\chi=\chi_1$ (traversed by mode~1),  and one channel has transmittance $\eta={\eta_2}$ and noise $\chi=\chi_2$ (traversed by mode~2).
After traversing the channels the density operator of the output mixed state
can be calculated in the Fock basis through the use of Eq.~\eqref{k2}, giving (see Appendix for derivation),
\begin{eqnarray}\label{output}
{\rho _{out}} = \sum\limits_{a = 0}^\infty  {\sum\limits_{b = 0}^\infty  {\sum\limits_{c = 0}^\infty  {\sum\limits_{d = 0}^\infty  {{\rho _{abcd}}{{\left| a \right\rangle }_1}{{\left\langle c \right|}_1} \otimes {{\left| b \right\rangle }_2}{{\left\langle d \right|}_2}} } } } ,
\end{eqnarray}
with
\begin{eqnarray}\label{symn}
\begin{array}{l}
{\rho _{abcd}} = \left( {{{\left\langle b \right|}_2}{{\left\langle a \right|}_1}} \right){\rho _{out}}\left( {{{\left| c \right\rangle }_1}{{\left| d \right\rangle }_2}} \right) = \\
\\
\phi _1^{ - 2}\phi _2^{ - 2}\sum\limits_{j = 0}^{\min \{ b + \ell  - \ell ',d + \ell  - \ell '\} } {\sum\limits_{\ell ' = 0}^{\min \{ b,d\} } {\sum\limits_{\ell  = 0}^\infty  {{q_{b + \ell  - \ell '}}{q_{d + \ell  - \ell '}}} } } \\
\\
 \times \sqrt {{}^a{C_{a - b - \ell  + \ell ' + j}}{}^c{C_{c - d - \ell  + \ell ' + j}}{}^{b + \ell  - \ell '}{C_j}{}^{d + \ell  - \ell '}{C_j}} \\
\\
 \times {\left( {\phi _1^{ - 1}{\zeta _1}} \right)^{\left( {b + d + 2(\ell  - \ell ' - j)} \right)}}{\left( {1 - \phi _1^{ - 2}} \right)^{a - b - \ell  + \ell ' + j}}\\
\\
 \times {\left( {1 - \zeta _1^2} \right)^j}\sqrt {{}^b{C_{\ell '}}{}^d{C_{\ell '}}{}^{b + \ell  - \ell '}{C_\ell }{}^{d + \ell  - \ell '}{C_\ell }} \\
\\
 \times {\left( {\phi _2^{ - 1}{\zeta _2}} \right)^{\left( {b + d - 2\ell '} \right)}}{\left( {1 - \phi _2^{ - 2}} \right)^{\ell '}}{\left( {1 - \zeta _2^2} \right)^\ell }
\end{array}
\end{eqnarray}
$$
{\rm if \ } a - b = c - d {\rm \, \, and \, \,  } {a - b - \ell  + \ell ' + j \ge 0},$$
$$
 {\rm otherwise \ } \rho_{abcd}=0.
$$
Here,
\begin{eqnarray}\label{Def}
\begin{array}{l}
{\phi _1} = \sqrt {1 + {\chi _1}/2} ,\,\,{\zeta _1} = {\eta _1}/\sqrt {1 + {\chi _1}/2} \\
\\
{\phi _2} = \sqrt {1 + {\chi _2}/2} ,\,\,{\zeta _2} = {\eta _2}/\sqrt {1 + {\chi _2}/2} .
\end{array}
\end{eqnarray}
In the case of a quantum limited attenuator channel (where ${\chi_1} = {\chi_2} = 0$), Eq.~\eqref{symn} takes the following form
\begin{eqnarray}\label{sym}
\begin{array}{l}
{\rho _{abcd}} = \sum\limits_{\ell  = \max \{ 0,a - b\} }^\infty  {{q_{b + \ell }}\,{q_{d + \ell }}} \\
\\
 \times \sqrt {{}^{b + \ell }{C_{b - a + \ell }}{}^{d + \ell }{C_{d - c + \ell }}} \,\,\,\eta _{_1}^{a + c}\,\,{\left( {1 - \eta _{_1}^2} \right)^{b - a + \ell }}\\
\\
 \times \sqrt {{}^{b + \ell }{C_\ell }{}^{d + \ell }{C_\ell }} \,\,\eta _2^{b + d}{\left( {1 - \eta _2^2} \right)^\ell }
\end{array}
\end{eqnarray}
if $a - b = c - d $, otherwise $\rho _{abcd} = 0$.

In the asymmetric setting mode~1 is kept at the sending station (and we set $\eta_1=1$ and $\chi_1=0$), while mode~2 is passed through the channel, and in this case 
the density operator of the output mixed state
can be described by Eq.~\eqref{output}, where $\rho _{abcd}$ is given by

\begin{eqnarray}\label{asymn}
\begin{array}{l}
{\rho _{abcd}} = \phi _2^{ - 2}{q_a}\,{q_c}\sum\limits_{\ell ' = 0}^{\min \left\{ {b,d} \right\}} {\sqrt {{}^b{C_{\ell '}}{}^d{C_{\ell '}}{}^a{C_{a - b + \ell '}}{}^c{C_{c - d + \ell '}}} } \\
\\
 \times \,{\left( {1 - \phi _2^{ - 2}} \right)^{\ell '}}{\left( {\phi _2^{ - 1}{\zeta _2}} \right)^{\left( {b + d - 2\ell '} \right)}}{\left( {1 - \zeta _2^2} \right)^{a - b + \ell '}}
\end{array}
\end{eqnarray}
if $a - b = c - d$ and $a - b + \ell ' \ge 0$, otherwise $\rho _{abcd} = 0$. In the quantum limited attenuator channel for the asymmetric case (${\chi_2} = 0$), Eq.~\eqref{asymn} takes the following form
\begin{eqnarray}\label{asym}
{\rho _{abcd}} = {q_a}\,{q_c}\sqrt {{}^a{C_{a - b}}{}^c{C_{c - d}}} {\left( {1 - \eta _{_2}^2} \right)^{a - b}}\eta _{_2}^{b + d}
\end{eqnarray}
if $a - b = c - d \ge 0$, otherwise $\rho _{abcd} = 0$.

\subsection{Determination of the Entanglement}
We adopt the logarithmic negativity in order to evaluate the entanglement since it gives an upper bound on the distillable entanglement \cite{LN}. The logarithmic negativity of a bipartite  state $\rho $ is defined as
${E_{LN}}(\rho ) = {\log _2}\left( {1 + 2N(\rho )} \right)$,
where $N(\rho )$ is the negativity defined as the absolute value of the sum of the negative eigenvalues of ${\rho ^{PT}}$, the partial transpose of $\rho $ with respect to either subsystem. For a pure entangled state in the form of $\left| \psi  \right\rangle  = \sum\limits_{n = 0}^\infty  {{q_n}} {\left| n \right\rangle _1}{\left| n \right\rangle _2}$, the logarithmic negativity can be calculated analytically as
${E_{LN}}(\left| \psi  \right\rangle ) = 2{\log _2}\left( {\sum\limits_{n = 0}^\infty  {{q_n}} } \right)$.
Hence, the logarithmic negativity of all our initial states (before propagating through the channel) can be calculated analytically (except the $\rm PSS_s$ state and the $\rm PAS_s$ state). However, in general it is not possible to analytically determine the logarithmic negativity of the states after transmission over the lossy channel.

In this work a numerical method is deployed for evaluation of the logarithmic negativity in the general case. We determine the logarithmic negativity of our evolved states via the use of artificial cutoffs in the size of the density (matrix) operators. In this sense our approach is similar to that adopted in \cite{2} for the case of  
noiseless loss channels. Relative to \cite{2}, we will need to introduce an additional cutoff term.

Consider an initially entangled state in the form of Eq.~\eqref{initial} passing through the channel.
The partial transpose of the evolved density operator ${\rho _{out}}$ in Eq.~\eqref{output}, with respect to mode~2, is given by
\begin{eqnarray}\label{pt1}
\rho _{out}^{PT} = \sum\limits_{a = 0}^\infty  {\sum\limits_{b = 0}^\infty  {\sum\limits_{c = 0}^\infty  {\sum\limits_{d = 0}^\infty  {{\rho _{adcb}}{{\left| a \right\rangle }_1}{{\left\langle c \right|}_1} \otimes {{\left| b \right\rangle }_2}{{\left\langle d \right|}_2}} } } } ,
\end{eqnarray}
where
${\rho _{adcb}} = \left( {{{\left\langle d \right|}_2}{{\left\langle a \right|}_1}} \right){\rho _{out}}\left( {{{\left| c \right\rangle }_1}{{\left| b \right\rangle }_2}} \right)$,
and where the elements $\rho _{adcb}$ are zero unless $a + b = c + d = F \ge 0$.
Note that here the partial transpose of the density operator is block diagonal in the Fock state basis, where the blocks correspond to $F = 0,1,2,...$ (corresponding to each $F$, there is a $\left( {F + 1} \right) \times \left( {F + 1} \right)$ block). For numerical computation of the logarithmic negativity, we
are required to approximate $\rho _{out}^{PT}$ by limiting its size, i.e. creating a truncated $\rho _{out}^{PT}$. In the symmetric setting, first we will set a cutoff on $F$, i.e. $\max \left( F \right) = {F_{\max }}$, and then we will set a cutoff on $\ell$ in Eq.~\eqref{symn} or Eq.~\eqref{sym}, i.e. $\max \left( \ell \right) = {\ell_{\max }}$. The value of ${F_{\max }}$ and ${\ell_{\max }}$  introduced should be large enough compared with the mean photon number of the state, and the trace of the truncated $\rho _{out}^{PT}$  ($\sim 1$) can be used as a measure of the validity of our chosen cutoff values.
 In our simulations in the next sections we will choose $F_{\max }=\ell_{\max }=10$ in the low squeezing regime (3dB) and $F_{\max }=\ell_{\max }=50$ in the high squeezing regime (10dB).
The logarithmic negativity can then be determined directly from the negative eigenvalues of the truncated $\rho _{out}^{PT}$.
For the asymmetric setting a similar exposition is utilized except we will only need to set a cutoff on $F$.

Note, that the density operators of the $\rm PSS_s$ state and the $\rm PAS_s$ state are not in the form of Eq.~\eqref{initial}, and therefore  Eqs.~\eqref{symn}-\eqref{asym} cannot be used directly to evolve them. However, a very similar approach to that described by Eqs.~\eqref{symn}-\eqref{asym} can be used to calculate the output density operator and the logarithmic negativity (the partial transpose of the output density operator is still block diagonal in the Fock state basis). Note also, that the density operator of a NOON state is different from the form of Eq.~\eqref{initial}. Hence, again we are not able to use Eqs.~\eqref{symn}-\eqref{asym} to calculate the evolved density operator. Instead, we utilize Eq.~\eqref{k2} to calculate the elements of the evolved state. However, in the case of noisy channels, the final density operator again possesses an infinite number of elements. Similar to before, we set a truncation cutoff, but this time on the $\ell '$ in Eq.~\eqref{k2}. From the
eigenvalues of the now truncated NOON density operator, the logarithmic negativity is once again determined.

\section{Results}

 Unless stated otherwise all calculations shown in this work assume an excess channel (Gaussian) noise  of $\chi=0.02$ (i.e. in symmetric setting $\chi_1=\chi_2=0.02$; in asymmetric setting $\chi_1=0, \chi_2=0.02$). We note a value of $\chi<0.02$ is consistent with receiver noise in modern detectors \cite{noiseref}. In the first instance we ignore any realistic operational constraints, and probe the evolution of our states under conditions where the initial entanglement of all states is equal. We could have  chosen other metrics  as the equality condition (e.g. energy, covariance matrix), but these first calculations  better demonstrate the important role played by the relative initial entanglement of the states,  and will prove  more useful when we come to discuss the impact of non-Gaussian operations on the TMSV states within anticipated operational settings.
\begin{figure}[!t]
    \begin{center}
   {\includegraphics[width=3 in, height=4.7 in]{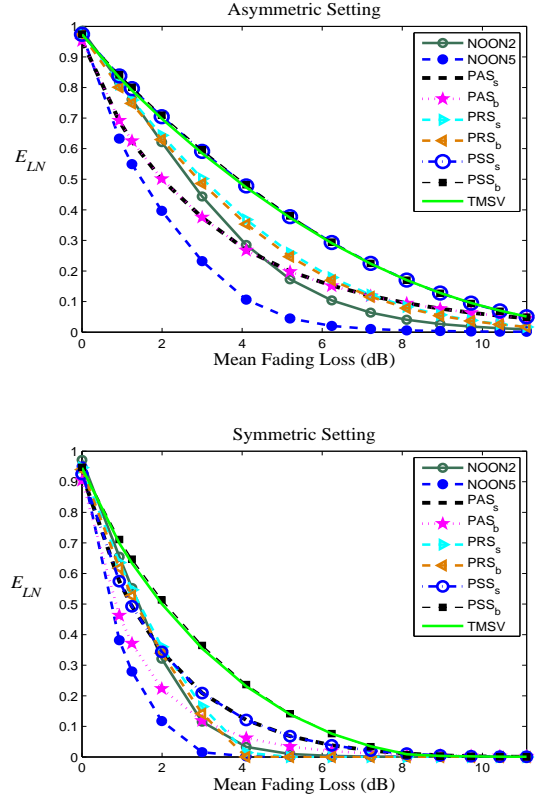}}
    \caption{Logarithmic negativity of the ensemble-average state resulting from asymmetric (top) and symmetric (bottom) settings with all states initially possessing ${E_{LN}=}$1~ebit.}\label{fig:0}
    \end{center}
\end{figure}

\begin{figure}[!t]
    \begin{center}
   {\includegraphics[width=3.05 in, height=2.1 in]{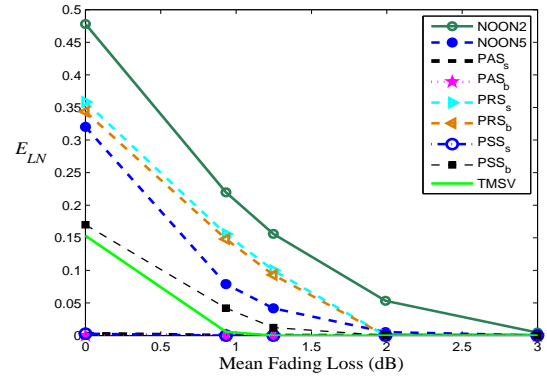}}
    \caption{Logarithmic negativity of the ensemble-average state in the symmetric setting with a large Gaussian noise of $\chi_1=\chi_2=0.4$. Again, all states initially possess ${E_{LN}=}$1~ebit.}\label{fig:3}
    \end{center}
\end{figure}

In Fig.~\ref{fig:0} the logarithmic negativity of the ensemble-average state is plotted as a function of channel loss. The top figure is for  asymmetric channels and the bottom figure is for symmetric channels (in this bottom figure the channel loss shown is applied to both channels - thus explaining the reduced entanglement). Here we assume all of our states have the same initial entanglement of 1 ebit (${E_{LN}}=1$) before transmission (achieved for some states by using differently squeezed initial TMSV states and applying non-Gaussian operations to them). The TMSV state shown here has an initial squeezing of 3dB. Note that the pure NOON states, regardless of the value of $n$, contain 1 ebit of entanglement. The abscissa corresponds to $\int_0^{{\eta _0}} {{\eta ^2}p\left( \eta  \right)} \,\,d\eta $, and represents mean fading losses under different channels (different $\sigma_b$). The fading losses shown  cover the range anticipated for satellite-to-ground links (i.e. downlinks).

The main trend seen from Fig.~\ref{fig:0} is that in terms of retaining entanglement, the TMSV state shows almost the same robustness as the PSS state ($\rm PSS_s$ and $\rm PSS_b$ in the asymmetric setting, and $\rm PSS_b$ in the symmetric setting), with both these states showing more robustness than the other states.
Note in the symmetric setting single-mode photon subtraction and single-mode photon addition lead to the same entanglement robustness.
Although not shown, similar trends (albeit with less entanglement surviving) are found also for the higher-loss (uplink) channels. Taken at face value, Fig.~\ref{fig:0} implies there is effectively no
advantage in using non-Gaussian states through atmospheric fading channels if the initial entanglement of such states is equal to the TMSV state.

The impact of much higher Gaussian noise in the symmetric setting is investigated in Fig.~\ref{fig:3}. Here we have adopted a $\chi_1=\chi_2=0.4$ simply to highlight the effect large Gaussian noise can have on the relative robustness of the states. As we see, such a high noise level has a major impact on the entanglement trends, with in this case the TMSV state evolving to zero entanglement for very low losses, but with most of the other states retaining their entanglement relatively better. The most robust state shown here is the $n=2$ NOON state. Clearly, in these higher noise conditions, most of the non-Gaussian states perform better than the TMSV state.

We now investigate initial conditions more likely to be present in operational scenarios.
We consider a  scenario in which all the non-Gaussian operations are applied to identical TMSV states.
For each non-Gaussian operation we select an optimal value of $T$. The determination of the optimal $T$ is a function of the initial squeezing of the TMSV state, and also whether it is the initial $E_{LN}$ or $R_E$ that is being maximized. Optimizing $E_{LN}$ for PSS states and PAS states always leads to an optimal value of $T=1$, independent of the initial squeezing. $R_E$ brings in the creation probability of the initial state $P_c$ (where $P_c$ refers to the relevant $P_{sb}$, $P_{ss}$, etc. of Sec.~II) and therefore brings in an additional dependence on $T$, relative to $E_{LN}$.

\begin{figure}[!t]
    \begin{center}
   {\includegraphics[width=3 in, height=4.7 in]{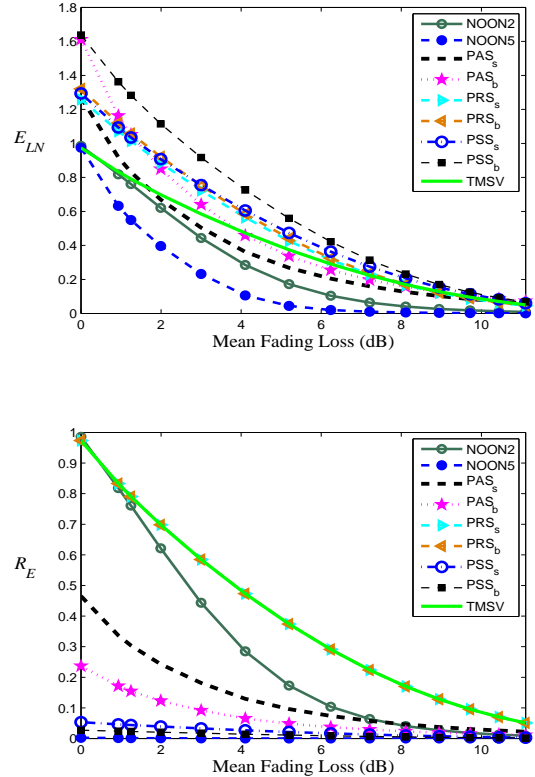}}
    \caption{The $E_{LN}$ (top) and $R_E$ (bottom) in the asymmetric setting  where the initial TMSV state had a squeezing of 3dB, this state being used for the non-Gaussian operations.}\label{fig:1}
    \end{center}
\end{figure}

\begin{figure}[!t]
    \begin{center}
   {\includegraphics[width=3 in, height=4.7 in]{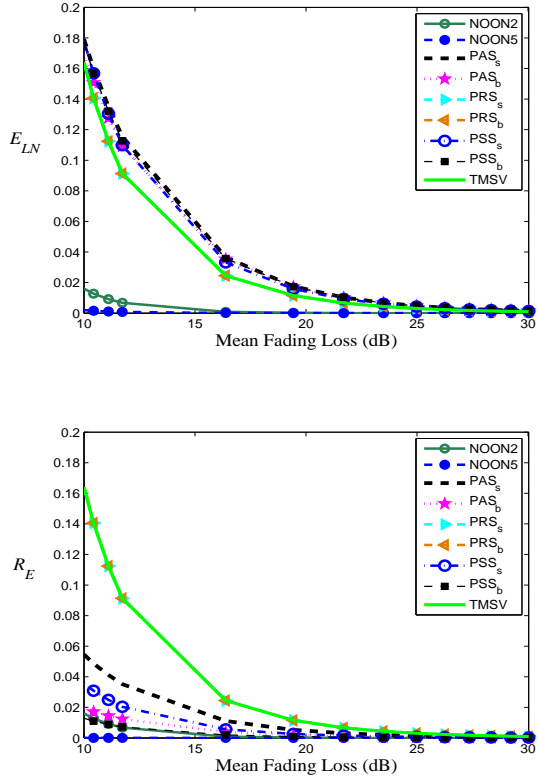}}
    \caption{The $E_{LN}$ (top) and $R_E$ (bottom) in the asymmetric setting where the initial TMSV state had a squeezing of 10dB, this state being used for the non-Gaussian operations.}\label{fig:2}
    \end{center}
\end{figure}

In Fig.~\ref{fig:1} the evolution of $E_{LN}$ (top) and $R_E$ (bottom) is shown for an initial TMSV state of 3dB squeezing. In Fig.~\ref{fig:2} the same results are shown for a squeezing of 10dB on the initial TMSV state.
Again, $E_{LN}$ and $R_E$ of the ensemble-average state are plotted as a function of the mean fading loss. The top figures of Figs.~\ref{fig:1}-\ref{fig:2} clearly show the benefit of applying the non-Gaussian operations to the TMSV states. Assuming that the sending rates (into the channel) for all the states are equal, then some non-negligible advantage of utilizing non-Gaussian states would be obtained. However, as can be seen from the $R_E$ values in the bottom plots of Figs.~\ref{fig:1}-\ref{fig:2}, once the probability of generation is taken into account any advantage gained from non-Gaussian states disappears. We can also see that single-mode non-Gaussian operations are more useful for the entanglement-generation rate than the two-mode operations - consistent with the former being produced with higher probability. Furthermore, the PAS states are more useful than the PSS states in terms of the entanglement-generation rate, again since the PAS states are created with higher probability. Note that as discussed earlier when $T=1$ the replacement operation has no effect on the initial TMSV state. However, optimizing $R_E$ for PRS states results in an optimal value of $T=1$. As such, the photon replacement operation is redundant in this context - the initial PRS state at the operating point $T=1$ is not a non-Gaussian state but rather just a TMSV state (other values of $T$ for PRS states may have benefits in other contexts). We also remind the reader that any non-unity probability associated with the generation of the single-photons utilized in the non-Gaussian operations is not accounted for in these plots of $R_E$ (e.g. heralded single-photon generation via a TMSV would entail an additional factor of  $\lambda^2(1-\lambda^2)$ per photon).

\begin{figure}[!t]
    \begin{center}
   {\includegraphics[width=3 in, height=4.7 in]{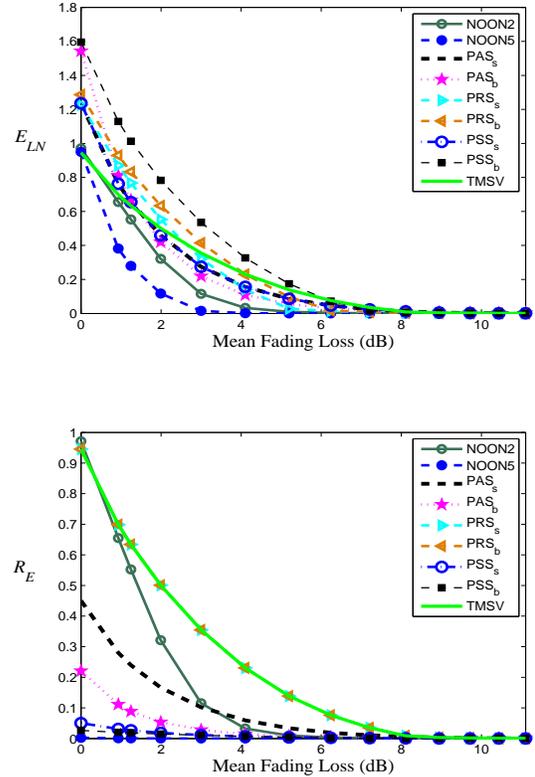}}
    \caption{The $E_{LN}$ (top) and $R_E$ (bottom) in the symmetric setting where the initial TMSV state had a squeezing of 3dB, this state being used for the non-Gaussian operations.}\label{fig:5}
    \end{center}
\end{figure}

A final result for this section is given in Fig.~\ref{fig:5} where we consider the symmetric scenario in which  an initial TMSV state of 3dB squeezing is utilized. The general trends we discussed above are again seen, albeit at lower entanglement values and rates (relative to the asymmetric setting) as a consequence of the equal mean fading loss in both channels.

Additional calculations beyond those illustrated in this section have been carried out,  covering the full spectrum of atmospheric channel and noise conditions anticipated to be relevant to LEO communications. All of these calculations result in  similar trends to those indicated above.

\section{Distillation, Memory and Single Photons}

Let us now assume a slightly modified scenario in which the variable transmittance is measured via the use of a separate coherent signal (e.g. a local oscillator  in
an orthogonal polarized mode to the signal sent through the channel). Although this adds some complexity to the system, the entanglement generated between stations will be enhanced. In this instance the TMSV states collected at the receiver during each transmittance window are Gaussian.
\begin{figure}[!t]
    \begin{center}
   {\includegraphics[width=3.05 in, height=2.1 in]{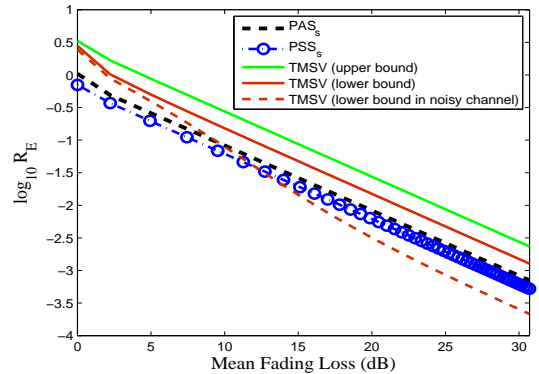}}
    \caption{Comparison of TMSV entanglement with the upper bound on non-Gaussian  distillable entanglement.}\label{fig:6}
    \end{center}
\end{figure}

Fig.~\ref{fig:6}  shows the values of $R_E$ in this scenario for the TMSV  (marked as TMSV (upper bound)), the PAS$_s$, and the PSS$_s$ states in the noiseless ($\chi_1=\chi_2=0$) asymmetric setting. Here a squeezing of 10dB on the initial TMSV state is adopted, and the logarithmic negativity is calculated as $E_{LN} = \int_0^{{\eta _0}} {p\left( \eta  \right)} \,E_{LN}\left( \eta  \right)\,d\eta $, where $E_{LN}\left( \eta  \right)$ is the logarithmic negativity of a state that has traversed a channel of transmittance $\eta$. The entanglement-generation rates shown are then $R_E={P_c}E_{LN}$, where $P_c$ refers to the relevant $P_{as}$ and $P_{ss}$ of Sec.~II ($P_c=1$ for TMSV states).
The entanglement-generation rates are higher (factor 2) in this case relative to the ensemble-average case, and quantifies the gain achieved for the additional complexity of sending  a local oscillator through the channel.

There remains the question as to whether, given some optimized distillation procedure, the non-Gaussian states could produce a higher entanglement-generation rate than that produced by TMSV states.
However, we note that $E_{LN}$ is an upper bound on the distillable entanglement.\footnote{If the ensemble state is used a  different (weaker) bound is needed since logarithmic negativity is not a convex function. This different bound is given by $E_{LN} = {\log _2}\left[ {1 + 2\int_0^{{\eta _0}} {p\left( \eta  \right)} \,{N}\left( \eta  \right)\,d\eta } \right]$ where ${{N}\left( \eta  \right)}$  is the negativity of a state that has traversed a channel of transmittance $\eta$.}
We also note that the conditional entropy  provides a lower bound on the distillable entanglement \cite{CE2,CE}.
The conditional entropy is calculated as ${E_{CE}} = \int_0^{{\eta _0}} {p\left( \eta  \right)} \,{E_{CE}}\left( \eta  \right)\,d\eta $, where ${E_{CE}}\left( \eta  \right)$ is the conditional entropy of a state that has traversed a channel of transmittance $\eta$. ${E_{CE}}\left( \eta  \right)$ is given by ${E_{CE}}\left( \eta  \right) = S({\rho _1}) - S(\rho )$, where $S(.)$ is the von Neumann entropy, $\rho $ is the density operator of the state after each realization of $\eta$, and ${\rho _1}$ refers to its reduced density operator with respect to the mode~1.
The  rate for the TMSV state (marked as TMSV (lower bound)) based on the conditional entropy is also shown in Fig.~\ref{fig:6}. Since the TMSV state after each realization of $\eta$ is still Gaussian, its conditional entropy can be calculated from its covariance matrix. For a given covariance matrix in the form of $M = \left\{ {A,C;{C^T},B} \right\}$,
the entropies can be calculated as $S({\rho _1}) = f(\sqrt {\det (A)} )$ and $S(\rho ) = f({\nu _1}) + f({\nu _2})$, where $f(x) = \frac{{x + 1}}{2}{\log _2}\left( {\frac{{x + 1}}{2}} \right) - \frac{{x - 1}}{2}{\log _2}\left( {\frac{{x - 1}}{2}} \right)$, and  ${\nu _{1,2}} = \left( {{{ ( {\Delta  \pm \sqrt {{\Delta ^2} - 4\det (M)} }  )} \mathord{\left/
 {\vphantom {{\left( {\Delta  \pm \sqrt {{\Delta ^2} - 4\det (\sigma_{CM})} } \right)} 2}} \right.
 \kern-\nulldelimiterspace} 2}} \right)^{1/2}$,  where $\Delta  = \det (A) + \det (B) + 2\det (C)$.

We can see from Fig.~\ref{fig:6} that distillation on the non-Gaussian states shown will not lead to any improvement in the entanglement-generation rates relative to those from the TMSV state (the distillable entanglement lower bound on the TMSV state is greater than the upper bounds on non-Gaussian states). This same conclusion is reached for all other non-Gaussian states studied. In fact, $R_E$ of the other non-Gaussian states studied is below that of the non-Gaussian states shown in Fig.~\ref{fig:6}. This figure also shows the TMSV rate based on its conditional entropy  in the case of a noisy asymmetric channel with $\chi_1=0,\chi_2=0.02$. As we can see for losses greater than about 10dB  the distillable entanglement lower bound on the TMSV state falls below the upper limit on the non-Gaussian states. For a noisy asymmetric channel and initial squeezing of 10dB on the initial TMSV state, we find  $\chi_2$ must be less than approximately $0.01$ (an attainable limit) for our argument on no improvement from distillation to be valid across all channels losses we consider. A similar result would apply to the symmetric case.

Distillation can still play a role in enhancing the quality (entanglement) of a received state relative to direct transmission of a TMSV state. However,  the results of Fig.~\ref{fig:6} (even though they apply only strictly in the infinite limit of state number) do imply that for optimizing  \emph{the number of received pulses above a specific entanglement target} in a low-noise channel, direct transmission of modestly squeezed TMSV states would be superior (the more squeezing the better).

Of course this last implication could be altered if a generation rate of non-Gaussian states beyond that adopted here could be achieved. One route to this would be through the availability of a long-term quantum memory coupled to a form of classical post-selection at the receiver. In such a scenario the receiver will feedback to the sender a transmittance measurement of the channel - possible due to the long coherence time of the channel (in the milliseconds range). On receipt of this classical message, the sending station will make a decision on whether to send the non-Gaussian states previously stored in memory (in addition to those being produced from the TMSV state)  dependent on the transmittance measurement being above some threshold $\eta_{th}$.
When below this threshold the sending station stores all non-Gaussian states produced from the TMSV state. Consider the following relation for some atmospheric channel and for some non-Gaussian state created from a squeezed TMSV,
\begin{eqnarray}\label{rob1v}
\frac{{{P_c}\int_{{\eta _{th}}}^{\eta _0}  {p(\eta )} E_{LN}^{ng}(\eta )d\eta }}{{\int_0^{\eta _0}  {p(\eta )} E_{LN}^g(\eta )d\eta }} = \mu  ,
\end{eqnarray}
where $\mu={\rm{Probability}}\left[ {\eta  \ge {\eta _{th}}} \right]$, and  $E_{LN}^{ng \ (g)}$ is the logarithmic negativity of the non-Gaussian (Gaussian) state that has traversed a channel of transmittance $\eta$. If there existed a non-Gaussian state that satisfied Eq.~\eqref{rob1v} for some $\eta_{th}>0$, then a quantum memory that could store the non-Gaussian state for a timescale of order $\Delta t_s(\mu^{-1}-1)$, where $\Delta t_s$ is the channel coherence timescale, would provide equality of the entanglement rates produced by the Gaussian and non-Gaussian states. However,  for the non-Gaussian states we have adopted (and their associated $P_c$ values), we find no solution to Eq.~\eqref{rob1v}  for some $\eta_{th}>0$ in any realistic channel settings (and useful initial squeezing values). Further, in this discussion we have ignored gains to be had by also storing the TMSV state. As such, a more likely use of quantum memory in the context of high-loss atmospheric channels would be in situations where both types of states are stored in memory so as to produce a true `on-demand' system. Assuming only output from the quantum memory is used for transmission, the entanglement rate generated between  stations (by any state) would simply increase linearly with the timescale of the memory.

\begin{figure}[!t]
    \begin{center}
   {\includegraphics[width=3 in, height=4.8 in]{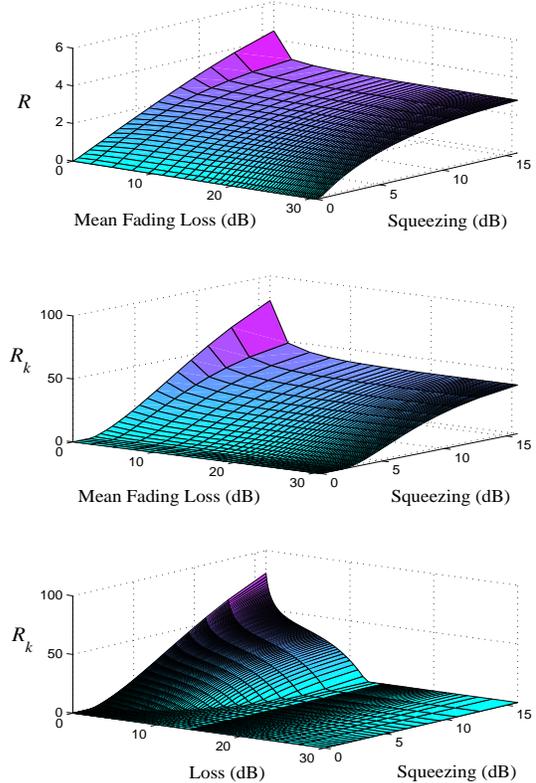}}
    \caption{A comparison of CV and discrete systems. The entanglement ratio $R$ (top) and the  key-rate ratio $R_k$ (middle) in the asymmetric setting with respect to the initial squeezing of the TMSV state, and the mean loss of the fading channel. The ratio $R_k$ (bottom) in the asymmetric setting with respect to the initial squeezing of the TMSV state, but for fixed channel loss. Note that, in the middle figure the CV key rate is 0.07 (bits per initial pulse) at the mean fading loss of 11dB and squeezing of 9.5dB.}\label{fig:ratio}
    \end{center}
\end{figure}

Finally, it is perhaps worth considering our results in relation to the entanglement generation-rate over atmospheric channels achievable by the transfer of single photons (quantum communication with LEO using single-photon technology has been achieved recently \cite{s11}). In the single-photon setting the  loss manifests itself directly as a reduced detection rate.  Fig.~\ref{fig:ratio} (top) shows $R$, the ratio of the entanglement-generation rate, $R_E$, achieved by asymmetric transfer of the TMSV state (our state with the highest $R_E$) to that achieved by single photon transfer. Here we have assumed the channel transmittance is measured, the single photon is the entangled partner of a Bell pair, and the rate of TMSV generation is equal to the rate of Bell-pair generation.  We have set $\chi_1=\chi_2=0.02$ for the CV entanglement determination, but have ignored any additional noise terms in the single-photon detectors or any reduction in the Bell pair entanglement caused by secondary channel effects (e.g. depolarization). We have also assumed a perfect Bell pair source (i.e. only pure two-photon Bell states produced). The ratio of the entanglement rates are shown as a function of the initial squeezing of the TMSV state, and as a function of the fading channel loss. From this  idealized setting for the single-photon production, we can see that above approximately 3dB squeezing (current state-of-the-art is $10$dB \cite{sque10}) the entanglement in the TMSV states dominates for all fading losses studied.

The comparison in Fig.~\ref{fig:ratio} (top) is of limited value in its own right for several reasons, most important of which is the fact that the forms of entanglement being compared  are fundamentally different. A true comparison of the merits of CV  versus Bell-pair entanglement technologies is complex and ultimately requires performance evaluation in the context of some operational measure (indeed operational measures are useful even in comparison amongst CV states - entanglement is not always a unique measure of operational efficiency). An important operational measure  would be the key rates of QKD.

The results shown in the middle figure of Fig.~\ref{fig:ratio}  for the same asymmetric setting illustrate the ratio, defined as $R_k$, of the key rates of a CV entanglement-based QKD scheme relative to a discrete entanglement-based  QKD scheme.
For the CV case, we have adopted the secret key rate $K_{CV}$ (bits per initial pulse) of a reverse reconciliation scheme with homodyne detection by the sender and receiver. This is derived as ${K_{CV}} = \int_0^{{\eta _0}} {p\left( \eta  \right)} \,{K_{CV}}\left( \eta  \right)\,d\eta$, where ${K_{CV}}\left( \eta  \right)$ is the key rate resulting from a TMSV state that has traversed a channel of transmittance $\eta$ (we do not investigate key rates generated by our non-Gaussian states due to their low production rates in realistic operational settings). Since the TMSV state after each realization of $\eta$ is still Gaussian, the CV key rate can be calculated through the use of Eqs.~(5)-(11) of \cite{neda2}. Again, we have set $\chi_1=\chi_2=0.02$ for the CV system. The discrete QKD scheme adopted here is the entanglement-based QKD of \cite{lo1}, in which a source emits a state described by Eq.~(2) of \cite{lo1} (at the same generation rate of the TMSV state) which is then transmitted through the fading channel to the receiver. The secret key rate $K_{DV}$ (bits per initial pulse) can be derived as ${K_{DV}} = \int_0^{{\eta _0}} {p\left( \eta  \right)} \,{K_{DV}}\left( \eta  \right)\,d\eta$, where ${K_{DV}}\left( \eta  \right)$ is the key rate resulting from a state (described by Eq.~(2) of \cite{lo1}) that has traversed a channel of transmittance $\eta$. ${K_{DV}}\left( \eta  \right)$ is determined through the use of Eqs.~(9)-(12) of \cite{lo1}, with the variables defined in \cite{lo1} set as $q=0.5$, $f(E_\mu)=1.22$, $Y_{0A}=Y_{0B}=6.024 \times {10^{ - 6}}$, $e_0=0.5$, $e_d=0.015$, and a value of $\mu=0.175$.\footnote{
 This value of $\mu$ is effectively optimal. If put to a more practical setting of say $\mu=0.05$ (as discussed in \cite{lo1}) the probability of producing a single two photon Bell states is significantly reduced, and the ratio $R_k$ would then be about twice that shown.} The ratio $R_k$ is then given by ${K_{CV}}/{K_{DV}}$. Note, this ratio implicitly assumes the number of exchanges between sender and receiver are infinite. Note also, in order to largely remove detector issues from the comparison we have set all detector efficiencies to $1$ (the efficiency of the homodyne detectors will likely be larger in practice).

As can be seen from  this middle figure,
significant enhancements in CV QKD  relative to discrete  QKD  are present. Of course, any comparison of CV QKD and discrete QKD is ultimately more complex than that provided here. Quite different detectors are deployed in the different schemes (different efficiencies, dark counts, etc.) and variants on the implementation strategies (and assumptions adopted) of both schemes are available which can impact final key rates significantly. Nonetheless, the results of the entanglement-based QKD comparison shown in Fig.~\ref{fig:ratio} (middle) are indicative of the intrinsic advantage infinite-dimension Hilbert space systems possess, and show that this advantage can persist over the fading channels anticipated for communications between terrestrial stations and low-earth orbit satellites.

In the bottom figure of Fig.~\ref{fig:ratio} we repeat our comparison of the quantum key rates just described, but this time for \emph{fixed} channel losses. The dramatic differences seen here relative to the middle figure clearly demonstrates the impact the fading channel can have on a quantum communication outcome - in this case on the quantum key rates of an entanglement-based CV system relative to those of an entanglement-based qubit system.


\section{Conclusion}

In this work we have explored the entanglement robustness of a wide range of non-Gaussian states against decoherence under atmospheric fading channels.
We have found that if the sending rates of all the states could be equalized, e.g. via an on-demand system derived from quantum memory, some non-Gaussian states can be produced and used to provide an entanglement transfer advantage relative to the usage of Gaussian states. If, however, the non-Gaussian states are produced (and sent) just-in-time via non-Gaussian operations on arriving TMSV states, then simply sending the arriving TMSV state over the atmospheric channel would most likely be the best option in terms of the entanglement-generation rate. The calculations presented here should be of value in assessments of the different technology solutions under consideration for future space-based quantum communications.


\appendix
Let us consider an initial entangled state with density operator in the form of Eq.~\eqref{initial}.
Considering the symmetric setting, mode~1 (2) evolves according to Eq.~\eqref{k2} with transmittance ${\eta _1}$  (${\eta _2}$) and noise ${\chi_1}$ (${\chi_2}$).
The final density operator of the output state
can be calculated through the use of Eq.~\eqref{k2}, giving
\begin{eqnarray}\label{App1}
\begin{array}{l}
{\rho _{out}} = \sum\limits_{m = 0}^\infty  {\sum\limits_{n = 0}^\infty  {{q_m}{q_n}} }  \times \\
\\
\phi _1^{ - 2}\sum\limits_{j' = 0}^\infty  {\sum\limits_{j = 0}^{\min \{ m,n\} } {\sqrt {{}^{m - j + j'}{C_{j'}}{}^{n - j + j'}{C_{j'}}{}^m{C_j}{}^n{C_j}} } } \\
\\
 \times {\left( {\phi _1^{ - 1}{\zeta _1}} \right)^{\left( {m + n - 2j} \right)}}{\left( {1 - \phi _1^{ - 2}} \right)^{j'}}{\left( {1 - \zeta _1^2} \right)^j}\\
\\
 \times {\left| {m - j + j'} \right\rangle _1}{\left\langle {n - j + j'} \right|_1} \otimes \\
\\
\phi _2^{ - 2}\sum\limits_{\ell ' = 0}^\infty  {\sum\limits_{\ell  = 0}^{\min \{ m,n\} } {\sqrt {{}^{m - \ell  + \ell '}{C_{\ell '}}{}^{n - \ell  + \ell '}{C_{\ell '}}{}^m{C_\ell }{}^n{C_\ell }} } } \\
\\
 \times {\left( {\phi _2^{ - 1}{\zeta _2}} \right)^{\left( {m + n - 2\ell } \right)}}{\left( {1 - \phi _2^{ - 2}} \right)^{\ell '}}{\left( {1 - \zeta _2^2} \right)^\ell }\\
\\
 \times {\left| {m - \ell  + \ell '} \right\rangle _2}{\left\langle {n - \ell  + \ell '} \right|_2}.
\end{array}
\end{eqnarray}
Now we will rewrite the above density operator in the following form
\begin{eqnarray}\label{App2}
{\rho _{out}} = \sum\limits_{a = 0}^\infty  {\sum\limits_{b = 0}^\infty  {\sum\limits_{c = 0}^\infty  {\sum\limits_{d = 0}^\infty  {{\rho _{abcd}}{{\left| a \right\rangle }_1}{{\left\langle c \right|}_1} \otimes {{\left| b \right\rangle }_2}{{\left\langle d \right|}_2}} } } } ,
\end{eqnarray}
where
$
{\rho _{abcd}} = \left( {{{\left\langle b \right|}_2}{{\left\langle a \right|}_1}} \right){\rho _{out}}\left( {{{\left| c \right\rangle }_1}{{\left| d \right\rangle }_2}} \right).
$
Comparing Eq.~\eqref{App1} and Eq.~\eqref{App2} we  have, $a = m - j + j',\,\,c = n - j + j', \
b = m - \ell  + \ell ',\,\,$ and $d = n - \ell  + \ell '$;
which gives,
$m = b + \ell  - \ell ',\,\,n = d + \ell  - \ell ' \ $ and
$j' = a - b - \ell  + \ell ' + j$.
Substituting $m, n, j'$ into Eq.~\eqref{App1} gives,
\begin{eqnarray}\label{App6}
\begin{array}{l}
{\rho _{out}} = \sum\limits_{a = 0}^\infty  {\sum\limits_{b = 0}^\infty  {\sum\limits_{c = 0}^\infty  {\sum\limits_{d = 0}^\infty  {\sum\limits_{\ell ' = 0}^{\min \{ b,d\} } {\sum\limits_{\ell  = 0}^\infty  {{q_{b + \ell  - \ell '}}} } {q_{d + \ell  - \ell '}}} } } } \phi _1^{ - 2} \times \\
\\
\sum\limits_{j' = 0}^{\min \{ a,c\} } {\sum\limits_{j = 0}^{\min \{ b + \ell  - \ell ',d + \ell  - \ell '\} } {\sqrt {{}^a{C_{j'}}{}^c{C_{j'}}{}^{b + \ell  - \ell '}{C_j}{}^{d + \ell  - \ell '}{C_j}} } } \\
\\
 \times {\left( {\phi _1^{ - 1}{\zeta _1}} \right)^{\left( {b + d + 2(\ell  - \ell ' - j)} \right)}}{\left( {1 - \phi _1^{ - 2}} \right)^{j'}}{\left( {1 - \zeta _1^2} \right)^j}\\
\\
 \times {\delta _{j',(a - b - \ell  + \ell ' + j)}}{\left| a \right\rangle _1}{\left\langle b \right|_1} \otimes \\
\\
\phi _2^{ - 2}\sqrt {{}^b{C_{\ell '}}{}^d{C_{\ell '}}{}^{b + \ell  - \ell '}{C_\ell }{}^{d + \ell  - \ell '}{C_\ell }} \\
\\
 \times {\left( {\phi _2^{ - 1}{\zeta _2}} \right)^{\left( {b + d - 2\ell '} \right)}}{\left( {1 - \phi _2^{ - 2}} \right)^{\ell '}}{\left( {1 - \zeta _2^2} \right)^\ell }\\
\\
 \times {\left| b \right\rangle _2}{\left\langle d \right|_2},
\end{array}
\end{eqnarray}
where $\delta _{i,j}$ is the Kronecker delta function.
From Eq.~\eqref{App6}, the matrix elements $\rho _{abcd}$ of density operator $\rho_{out}$ can be written in the form of Eq.~\eqref{symn}.

We can also use the density operator in Eq.~\eqref{App1} directly to compute the logarithmic negativity.
The partial transpose, $\rho _{out}^{PT}$, of the density operator with respect to mode~2 is given by Eq.~\eqref{App1} except in the last term we have the substitution,
$$
{\left| {m - \ell  + \ell '} \right\rangle _2}{\left\langle {n - \ell  + \ell '} \right|_2} \rightarrow
{\left| {n - \ell  + \ell '} \right\rangle _2}{\left\langle {m - \ell  + \ell '} \right|_2}.
$$
For numerical computation of the logarithmic negativity, we need to approximate $\rho _{out}^{PT}$ by limiting its size based on a cutoff on the variables of  $m,n,j'$ and $\ell '$, i.e. creating a truncated $\rho _{out}^{PT}$. We can first set a cutoff on $m$ and $n$, i.e. $\max \left( m \right) = {m_{\max }}$ and $\max \left( n \right) = {n_{\max }}$. By setting these two cutoffs we limit the size of the initial density operator, which means the number of elements contributing to each element of $\rho _{out}^{PT}$ is limited. Now we set  $F = m - j + j'+n - \ell  + \ell '$.
From this we see the maximum value of $j'$ and $\ell '$ is $F$, hence, instead of setting two cutoffs on $j'$ and $\ell '$, we can only set a cutoff on $F$, i.e. $\max \left( F \right) = {F_{\max }}$. By setting this cutoff we limit the number of photons which are produced by the noise in either of the modes. A final check on the truncated matrix (the cutoff values) is that its trace is very close to 1 (more formally we can set an accuracy parameter $\epsilon$ and  ensure the trace of the truncated matrix is between $1-\epsilon$ and $1$). For the asymmetric setting a similar exposition can be utilized except the variables $j$ and $j'$ are not required.



\begin{thebibliography}{1}

\bibitem{s1} M. Aspelmeyer et al., Long-distance quantum communication with entangled photons using satellites, IEEE Journal of Selected Topics in Quantum Electronics 9, 1541 (2003).



\bibitem{s4} R. Ursin et al., Space-quest, experiments with quantum entanglement in space, Europhysics News 40, 26 (2009).

\bibitem{s5} A.  C. Bonatoa,  V. Da Deppoa, G. Nalettoa,  P. Villoresia,  Link budget and background noise for satellite quantum key distribution, Adv. Space Res. 47, 802 (2011).


\bibitem{s7} J.-P. Bourgoin et al., A comprehensive design and performance analysis of low-earth orbit satellite quantum communication, New J. Phys. 15, 023006 (2013).



\bibitem{s10} J. Yin et al., Experimental quasi-single-photon transmission from satellite to Earth, Opt. Express 21, 20032 (2013).


\bibitem{s11} G. Vallone, D. Bacco, D. Dequal, S. Gaiarin, V. Luceri, G. Bianco, and P. Villoresi, Experimental satellite quantum communications, Phys. Rev. Lett. 115, 040502  (2015).

\bibitem{fso} L. C. Andrews and R. L. Phillips, \emph{Laser Beam Propagation Through Random Media}, 2nd ed. (SPIE, Bellingham, WA, 2005), Vol. PM152.

\bibitem{Dong} R. Dong, M. Lassen, J. Heersink, Ch. Marquardt, R. Filip, G. Leuchs, and U. L. Andersen, Continuous-variable entanglement distillation of non-Gaussian mixed states, Phys. Rev. A 82, 012312 (2010).

\bibitem{Usenko} V. C. Usenko, B. Heim, C. Peuntinger, C. Wittmann, C. Marquardt, G. Leuchs, and R. Filip, Entanglement of Gaussian states and the applicability to quantum key distribution over fading channels, New J. Phys. 14, 093048 (2012).

\bibitem{Heim} B. Heim, C. Peuntinger, N. Killoran, I. Khan, C. Wittmann, Ch. Marquardt, and G. Leuchs, Atmospheric continuous-variable quantum communication, New J. Phys. 16, 113018 (2014).

\bibitem{1} N. Hosseinidehaj and R. Malaney, Gaussian entanglement distribution via satellite, Phys. Rev. A 91, 022304 (2015).

\bibitem{Gaussian} C. Weedbrook, S. Pirandola, R. Garcia-Patron, N. J. Cerf, T. C. Ralph, J. H. Shapiro, and S. Lloyd, Gaussian quantum information, Rev. Mod. Phys. 84, 621 (2012).

\bibitem{1st_PSS} T. Opatrny, G. Kurizki, and D.-G. Welsch, Improvement on teleportation of continuous variables by photon subtraction via conditional measurement, Phys. Rev. A 61, 032302 (2000).

\bibitem{2} A. Kitagawa, M. Takeoka, M. Sasaki, and A. Chefles, Entanglement evaluation of non-Gaussian states generated by photon subtraction from squeezed states, Phys. Rev. A 73, 042310 (2006).

\bibitem{1st_PAS} F. Dell'Anno, S. De Siena, L. Albano, and F. Illuminati, Continuous-variable quantum teleportation with non-Gaussian resources, Phys. Rev. A 76, 022301 (2007).

\bibitem{telep} Y. Yang and F-L. Li, Entanglement properties of non-Gaussian resources generated via photon subtraction and addition and continuous-variable quantum-teleportation improvement, Phys. Rev. A 80, 022315 (2009).

\bibitem{3} S. L. Zhang, and P. van Loock, Distillation of mixed-state continuous-variable entanglement by photon subtraction, Phys. Rev. A 82, 062316 (2010).

\bibitem{4}	M. Allegra, P. Giorda, and M. G. A. Paris, Role of initial entanglement and non-Gaussianity in the decoherence of photon-number entangled states evolving in a noisy channel, Phys. Rev. Lett. 105, 100503 (2010).

\bibitem{Amp1} G. S. Agarwal, S. Chaturvedi, and A. Rai, Amplification of maximally-path-entangled number states, Phys. Rev. A 81, 043843 (2010).

\bibitem{Amp2} H. Nha, G. J. Milburn, and H. J. Carmichael, Linear amplification and quantum cloning for non-Gaussian continuous variables, New J. Phys. 12, 103010 (2010).

\bibitem{5}	G. Adesso, Simple proof of the robustness of Gaussian entanglement in bosonic noisy channels, Phys. Rev. A 83, 024301 (2011).

\bibitem{6}	K. K. Sabapathy, J. Solomon Ivan, and R. Simon, Robustness of non-Gaussian entanglement against noisy amplifier and attenuator environments, Phys. Rev. Lett. 107, 130501 (2011).

\bibitem{ME} J. Lee and H. Nha, Entanglement distillation for continuous variables in a thermal environment: Effectiveness of a non-Gaussian operation, Phys. Rev. A 87, 032307 (2013).

\bibitem{8}	S. N. Filippov, and M. Ziman, Entanglement sensitivity to signal attenuation and amplification, Phys. Rev. A 90, 010301(R) (2014).

\bibitem{9}	K. P. Seshadreesan, J. P. Dowling, and G. S. Agarwal, Non-Gaussian entangled states and quantum teleportation of Schrodinger-cat states, Phys. Scr. 90, 074029 (2015).

\bibitem{Oxford} T. J. Bartley and I. A. Walmsley, Directly comparing entanglement-enhancing non-Gaussian operations, New J. Phys. 17, 023038 (2015).

\bibitem{no_go1} J. Eisert, S. Scheel, and M. B. Plenio, Distilling Gaussian states with Gaussian operations is impossible, Phys. Rev. Lett. 89, 137903 (2002).

\bibitem{no_go2} G. Giedke and J. I. Cirac, Characterization of Gaussian operations and distillation of Gaussian states, Phys. Rev. A 66, 032316 (2002).

\bibitem{Oxford2013} T. J. Bartley, P. J. D. Crowley, A. Datta, J. Nunn, L. Zhang, and I. Walmsley, Strategies for enhancing quantum entanglement by local photon subtraction, Phys. Rev. A 87, 022313 (2013).

\bibitem{beamwander} D. Yu. Vasylyev, A. A. Semenov, and W. Vogel, Toward global quantum communication: Beam wandering preserves nonclassicality, Phys. Rev. Lett. 108, 220501 (2012).

\bibitem{sem} A. A. Semenov, F. T\"{o}ppel, D. Yu. Vasylyev, H. V. Gomonay, and W. Vogel, Homodyne detection for atmosphere channels, Phys. Rev. A 85, 013826 (2012).

 \bibitem{adapt}A. J. Hashmi, A. A. Eftekhar, A. Adibi and F. Amoozegar, Analysis of adaptive optics-based telescope arrays in a deep-space inter-planetary optical communications link between Earth and Mars, Optics Comm. 333, 120 (2014).

\bibitem{7}	C. Navarrete-Benlloch, R. Garcia-Patron, J. H. Shapiro, and N. J. Cerf, Enhancing quantum entanglement by photon addition and subtraction, Phys. Rev. A 86, 012328 (2012).

\bibitem{added} S. L. Zhang, Y. Dong, X. Zou, B. Shi, and G. C. Guo, Continuous-variable-entanglement distillation with photon addition, Phys. Rev. A 88, 032324 (2013).

\bibitem{NOON0} J. Fiurasek, Conditional generation of N-photon entangled states of light, Phys. Rev. A 65, 053818 (2002).

\bibitem{NOON1} P. Kok, H. Lee, and J. P. Dowling, Creation of large-photon-number path entanglement conditioned on photodetection, Phys. Rev. A 65, 052104 (2002).

\bibitem{NOON2}  H. Cable and J. P. Dowling, Efficient generation of large number-path entanglement using only linear optics and feed-forward, Phys. Rev. Lett. 99, 163604 (2007).

\bibitem{exp1} H. Takahashi, J. S. Neergaard-Nielsen, M. Takeuchi, M. Takeoka, K. Hayasaka, A. Furusawa, and M. Sasaki, Entanglement distillation from Gaussian input states, Nature Photonics 4, 178 (2010).

\bibitem{exp2} Y. Kurochkin, A.S. Prasad, and A. I. Lvovsky, Distillation of the two-mode squeezed state, Phys. Rev. Lett. 112, 070402 (2014).

\bibitem{added_exp1} A. Zavatta, S. Viciani, and M. Bellini, Quantum-to-classical transition with single-photon-added coherent states of light, Science 306, 660 (2004).

\bibitem{added_exp2} A. Zavatta, V. Parigi, and M. Bellini, Experimental nonclassicality of single-photon-added thermal light states, Phys. Rev. A 75, 052106 (2007).

\bibitem{NOON2015} M. Bohmann, J. Sperling, and W. Vogel, Entanglement and phase properties of noisy NOON states, Phys. Rev. A 91, 042332 (2015).

\bibitem{NOON_exp2} I. Afek, O. Ambar, and Y. Silberberg, High-NOON states by mixing quantum and classical light, Science 328, 879 (2010).

\bibitem{NOON_exp1} Y. Israel, I. Afek, S. Rosen, O. Ambar, and Y. Silberberg, Experimental tomography of NOON states with large photon numbers, Phys. Rev. A 85, 022115 (2012).

\bibitem{10} J. Solomon Ivan, K. K. Sabapathy, and R. Simon, Operator-sum representation for bosonic Gaussian channels, Phys. Rev. A 84, 042311 (2011).

\bibitem{LN} G. Vidal and R. F. Werner, Computable measure of entanglement, Phys. Rev. A 65, 032314 (2002).

\bibitem{noiseref} G. Nocerino, D. Buono, A. Porzio, and S. Solimeno, Survival of continuous variable entanglement over long distances,  Phys. Scr. T153, 014049 (2013).

 \bibitem{CE2}   J. Eisert, M.M. Wolf, Gaussian quantum channels, quant-ph/0505151 (2005).

\bibitem{CE}  M. M. Wolf, G. Geidke, J. I. Cirac, Extremality of Gaussian quantum states, Phys. Rev. Lett. 96, 080502 (2006).

\bibitem{sque10} T. Eberle, V. H\"{a}ndchen, and R. Schnabel, Stable control of 10 dB two-mode
squeezed vacuum states of light, Opt. Expr. 21, 11546 (2013).


\bibitem{neda2}  N. Hosseinidehaj and R. Malaney, Quantum key distribution over combined
atmospheric fading channels, IEEE International Conference on Communications (ICC), 7413, London (2015).

\bibitem{lo1}  X. Ma, C. Fung, and H. Lo, Quantum key distribution with entangled photon sources, Phys.  A 76, 012307 (2007).




\end{thebibliography}
\end{document}